\documentstyle[12pt,epsfig,amssymb]{article}
\begin{document}
\def\be{\begin{equation}}
\def\ee{\end{equation}}
                         \def\bearr{\begin{eqnarray}}
                         \def\eearr{\end{eqnarray}}
\def\alpmz{\relax\ifmmode \alpha_s(M_Z)\else $\alpha_s(M_Z)$\fi\chkspace}
\def\z0{\relax\ifmmode Z^0 \else {$Z^0$} \fi\chkspace}
\def\ep{{e$^+$e$^-$}\chkspace}
\def\alp{\relax\ifmmode \alpha_s\else $\alpha_s$\fi\chkspace}
\def\ep{{e$^+$e$^-$}}
\def\ggx{{$\gamma\gamma$}}
\def\sigg{{$\sigma_{\gamma\gamma}^{tot}$}}
\def\fg{{$F^{\gamma}_2$}}
\def\xg{{$x_{\gamma}$}}
\def\syy{{$\sqrt{s_{\gamma\gamma}}$}}
\def\gstar{{$\gamma^*\gamma^*$}}
\def\be{\begin{equation}}
\def\ee{\end{equation}}
\def\benum{\begin{enumerate}}
\def\eenum{\end{enumerate}}
\def\bitem{\begin{itemize}}
\def\eitem{\end{itemize}}
                         \def\eg{ {\em e.g.}~}
                         \def\etal{ {\em et al.}~}
                         \def\ie{ {\em i.e.}~}
                         \def\viz{ {\em viz.}~}

\def\go{\rightarrow}
\def\goes{\longrightarrow}
\def\hrar{\hookrightarrow}
\def\bul{\bullet}
\def\eplem{\mbox{$e^+e^-$}}
\def\gamgam{\mbox{$\gamma \gamma$}}
\def\gamp{\mbox{$\gamma p$}}
\def\rts{\mbox{$\sqrt{s}$}}
\def\ptmin{\mbox{$p_{\rm tmin}$}}
\def\siggmjet{\mbox{$\sigma (\gamma \gamma \rightarrow {\rm jets})$}}
\def\lsim{\:\raisebox{-0.5ex}{$\stackrel{\textstyle<}{\sim}$}\:}
\def\gsim{\:\raisebox{-0.5ex}{$\stackrel{\textstyle>}{\sim}$}\:}

\def\GG{$\gamma\gamma$}
\def\GE{$e\gamma$}
\def\eG{$e\gamma$}
\def\EPEM{$e^+e^-$}
\def\CPbar{\hbox{{\rm CP}\hskip-1.80em{/}}}

\begin{flushright}
hep-ph/0305071\\
DESY 03-054
\end{flushright}

\begin{center}

{\large\bf    
        Hadronic  Cross-sections in two photon  Processes at
        a Future Linear Collider\footnote{ E-mails: (a)rohini.godbole@desy.de, 
        (b) deroeck@mail.cern.ch, (c) igrau@ugr.es   
         (d) Giulia.Pancheri@lnf.infn.it} } \\
\vskip 25pt
{\bf                        R. M. Godbole$^a,$}
\footnote{Permanent Address: Centre for Theoretical Studies, Indian Institute of Science, Bangalore 560 012, India.}
\\
{\footnotesize\rm
Theory Division, DESY, Notkestrasse 85, D-22603 Hamburg, Germany}\\

\smallskip

{\bf  A. De Roeck$^b$} \\ 
{\footnotesize\rm 
                      CERN, CH-1211, Geneva-23, Switzerland}

\smallskip

{\bf                        A. Grau$^c$ } \\ 

{\footnotesize\rm 
                   CAFPE and DFTC, Universidad de Granada, Spain  }\\ 

\smallskip

{\bf                       G. Pancheri$^d$ } \\ 

{\footnotesize\rm 
                    Laboratori Nazionali di Frascati dell'INFN,  
                     Via E. Fermi 40, I 00044, Frascati, Italy.\\ } 
\vskip 15pt

{\bf                             Abstract 
}
\end{center}

\noindent
In this note we address the issue of measurability of the hadronic 
cross-sections at a future photon collider as well as for the 
two-photon processes at a future high energy linear $e^+e^-$ collider.  
We extend,  to  higher energy, our previous
estimates  of the accuracy  with which the \gamgam\ cross-section
needs to be measured, in order to distinguish  between different theoretical 
models of energy dependence  of the total cross-sections. 
We  show  that the necessary precision to discriminate
among these models is indeed possible at future linear colliders 
in the Photon Collider option. Further we note that even in the $e^+e^-$
option a measurement of the hadron production cross-section via \gamgam\ 
processes, with  an 
accuracy necessary to allow discrimination between different 
theoretical models, should be possible.
We also comment briefly on  the implications of these  predictions 
for hadronic backgrounds at the future TeV energy $e^+e^-$ collider CLIC.
\newpage
\section{ Introduction}
The rising total cross-section in proton-proton collisions was a very 
early indication of QCD processes at work, reflecting the fact that the
increasing energy allows a deeper probe of the structure of the colliding 
particles, leading to liberation of  more constituents which results in a 
higher scattering probability\cite{therise}. 
 The proton-proton and proton anti-proton cross-sections  are  
now  known experimentally to a very good precision over a large energy range. 
We still do not have a full theoretical understanding of these cross-sections 
starting from first principles, but there are various models of hadronic
interactions whose parameters can be completely fixed by the data and
which then allow for good predictions of the total cross-section in the 
high energy region, certainly up to LHC energies. 
Thus, although not everything is calculable from first principles in QCD, 
the total hadronic production at 
future hadronic accelerators  can be predicted.   Deviations from these
predictions, beyond the theoretical errors, could indicate   the  onset of
new physics, just like    the rise of total cross-section, first observed at
the ISR\cite{therise}, 
 was indeed the signal of hitherto undetected partonic interactions.

The situation is quite  different for the photon induced processes in that the
data cover a smaller energy range and also have larger errors.  
This renders the issue of measurement of  the total \gamgam\ cross-section 
at energies in the region  300-500 GeV,  very important both from 
the theoretical, as well as an experimental point of view. Indeed, the 
question  of hadron production in \gamgam\ collisions is interesting for 
achieving a good theoretical understanding of the rise of the hadronic 
cross-sections with energy, in the framework of QCD or otherwise, 
as well as from a much more pragmatic viewpoint of being able to
estimate the hadronic backgrounds~\cite{lcbkgd} at the next linear colliders,
particularly in the photon collider option as well in the $e^+e^-$ option
like CLIC~\cite{clic1} in the higher energy range. HERA and LEP have opened the 
way to an entire new field in QCD, the study of the hadronic interactions of 
the photon in terms of its  quark and gluon content~\cite{review}. The rise 
in the  total hadronic cross section begins to take place at centre of mass 
energies below 100 GeV but to determine the steepness of  the rise one needs 
points in the range 300-500 GeV and at even higher energies. These do not 
exist. Different models have been suggested in the context of rise of the 
hadronic cross-sections in $pp$ and $\bar p p$ processes. All of these 
`explain' the rise for the $pp$ and $\bar p p$ case equally well but 
differ substantially in their predictions for \gamgam\ collisions even at 
the modest values of the \gamgam\ energies that are currently available.
However, within various experimental uncertainities  they are all compatible 
with the current data.  Thus to  gain a good theoretical understanding of the 
total cross-section for \gamgam\ processes, as in the case of hadronic 
collisions, one needs much higher energies and better statistics than the one
currently available.

In the next section we discuss the  available data on
$\gamma\gamma$ total hadronic  cross sections at high energies. 
In section 3 we introduce  photon colliders and present recent model 
predictions for the cross-sections.  Section 4
details the simulation study made to assess the possible precision of the
experimental measurement of the hadronic cross-sections at the photon colliders.
In the next sections we discuss the required precision in the measurement 
of total hadronic cross-sections so as to distinguish among different models 
at the photon colliders and  $e^+e^-$ colliders respectively and then we 
end with conclusions. 

\section{Status of the currently available data and models} 

\begin{figure}[htb]
\centerline{
\includegraphics*[scale=0.45]{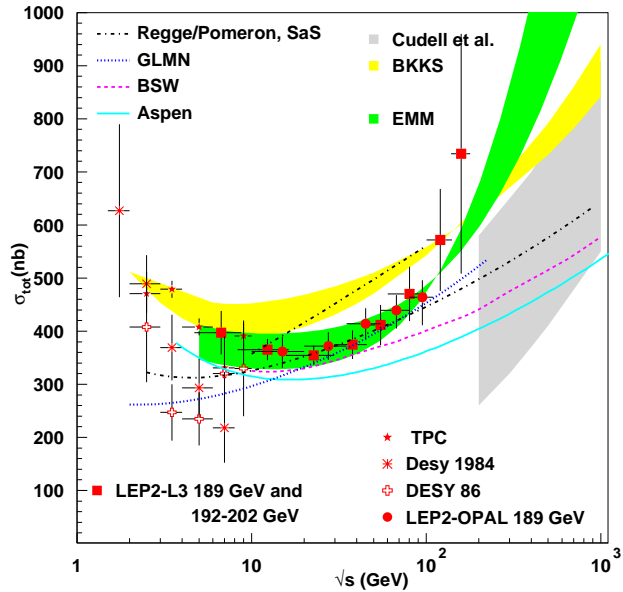}}
\caption{The predictions from factorization (proton like) 
models~\protect\cite{SAS,aspen,ttwu}
Regge-Pomeron exchange\protect\cite{glmn}  and a QCD structure function model
\protect\cite{BKS} together with those from the EMM\protect\cite{pancheri} are 
compared with the present data.}
\label{six}
\end{figure}

The currently available experimental information on total hadronic 
cross-sections  for photon induced processes comes from the 
$e^+e^-$ colliders PEP\cite{TPC}, PETRA\cite{DESY} and  LEP\cite{L3,OPAL}
 as well as HERA\cite{HERA}. LEP and HERA
provide the higher energy data. At LEP phase space limits the 
centre of mass system  (CMS) energy, $\sqrt{s_{\gamma\gamma}}$, 
of the \gamgam\ interactions to about 100 GeV, whereas at HERA 
$\sqrt{s_{\gamma p}}$ is higher.  However, at HERA the presence of the proton 
partly obscures the issue. 
Fig.~\ref{six} shows a collection of  data for the total hadronic
cross-section $\sigma (\gamma \gamma \rightarrow {\rm hadrons})$ from the
various $e^+e^-$ experiments in comparison with the predictions from a  
number of  theoretical models summarised in Ref. \cite{pancheri}. 
The predictions have been plotted from  ``proton-like'' models, 
labelled SaS\cite{SAS},
Aspen\cite{aspen},   BSW\cite{ttwu}, as well as from QCD and Regge inspired
models, like the curve labelled GLMN\cite{glmn} and the band labelled 
BKKS\cite{BKS}.  The band labelled EMM covers predictions of two different
formulations, inelastic and total. For the EMM, we have used two sets of
representative parameters\cite{pancheri}, both of which are obtained from the
$\gamma p$  cross-section following the procedure outlined in \cite{emmus}.
All models predict a rise of the cross-section with the collision 
energy \syy ,  but with very different slopes.  Also shown, by the region 
shaded in intermediate grey labelled Cudell et al,  is a
recent attempt of using only the low energy data to predict 
$\sigma_{\gamma p}$ and $\sigma_{\gamma \gamma}$ at TeV energies \cite{cudell}.
Within errors, their predictions are seen to cover
the range spanned by the predictions of almost all the models discussed here. 
Further  note that, until five years ago, the available data for
\gamgam\  processes stopped short of $\sqrt{s} = 20$ GeV and did not show 
any rise. The data in this energy range have very large errors, show a 
large spread and the compatibility of the different experiments is marginal. 
A re-measurement of this region, as planned at VLEP in Novisibirsk, may be very
useful even for understanding the high energy region as will be shown below.

The L3~\cite{L3} and OPAL~\cite{OPAL} data have drastically changed the 
situation.  Presently, the \gamgam\  cross-section data  indicate a very 
clear rise, which may be even stronger than in hadron-hadron collisions.
This can be shown through Regge inspired fits of the type
\begin{equation}
\sigma_{\gamgam}^{~\rm had} (s_{\gamgam}) = A~s_{\gamgam}^{\epsilon}  + 
B~ s_{\gamgam}^{-\eta}. 
\label{eq1}
\end{equation}
where $\epsilon $ and $\eta$ are expected to be process independent, 
i.e. the same in $pp$, $p{\bar p}$,  $\gamma p$ and $\gamma\gamma$ collisions.

First we note that the recent  data from LEP shown in
Fig.~\ref{six} have been corrected with models for 
acceptance and particularly for the invisible elastic cross sections and 
the low acceptance diffractive processes. To this end, 
Monte Carlo simulation programs like PYTHIA\cite{PYTHIA} and 
PHOJET\cite{phojet} have been used, which have different elastic and 
diffractive component predictions. The correction factors are different 
for these two and are large.  Both L3 and OPAL follow now the same strategy 
to present their data, by taking the average of the results obtained with 
both models. The data of the experiments agree in the region of overlap.

\begin{figure}[htb]
\centerline{
\includegraphics*[scale=0.45]{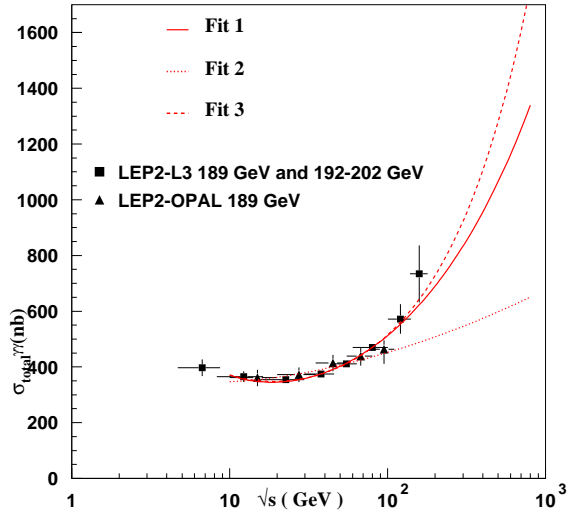}}
\caption{Data from OPAL and L3, shown with
combined statistical and systematic error apart from the
model dependence error, together with results from
fits. In Fit 1 all parameters of Eq. 1 are free. In Fit 2, $\epsilon$ is fixed
at the $pp/\bar p p $ value of 0.093, other parameters are free and
in Fit 3 a second pomeron term $C s^{\epsilon_1}$, with $\epsilon_1 = 0.418$ is added.}
\label{figfit}
\end{figure}

\begin{table}[htb]
\label{cross1}
\begin{center}
\caption{Results of fits to the OPAL and L3 total $\gamma\gamma$ cross 
sections, of the form $B s^{-\eta} + A s^{\epsilon}  + C s^{\epsilon_1} $.}
\vspace{0.3cm}
\begin{tabular}{|c||c|c|c|c|c||}
\hline
Data & $A$ (nb)& $B$ (nb)& $C$ (nb)& $\epsilon, \epsilon_1$ & $\chi^2$\\
\hline\hline
L3 & $47 \pm 14$ & $1154 \pm 158 $& -- & $\epsilon = 0.250 \pm 0.033$ & 2.4 \\
L3 & $187 \pm 4$ & $312 \pm 95$ & -- & $\epsilon = 0.093,$fixed  & 25 \\
L3 & $98 \pm 18$ & $958 \pm 162$ & $5.3\pm1.1$ & $\epsilon = 0.093$,fixed &  \\
&&&&$\epsilon_1 = 0.418$, fixed&1.3\\
L3+OPAL & $51 \pm 14$ & $1132 \pm 158$ & -- & $\epsilon = 0.240 \pm 0.032$ & 4.0 \\
L3+OPAL & $187 \pm 4$ & $310 \pm 91$ & -- &$\epsilon = 0.093$  fixed  & 26 \\
L3+OPAL & $103 \pm 18$ & $934 \pm 156$ & $5.0\pm 1.0$ &$\epsilon = 0.093$,fixed&\\
&&&&$\epsilon_1 = 0.418,$ fixed&2.8\\
\hline
\end{tabular}
\end{center}
\end{table}
Table 1 shows results of fits to the measurements
in the form of Eq.~\ref{eq1}, using the full errors (except for the model 
dependence), not taking into account correlations. Similar values for the 
parameters are obtained using only the statistical errors.  Since the power 
of the Regge term and its size cannot be both determined from the LEP data 
alone, we take the value measured in $pp$ and $\gamma p$ interactions 
as given in the PDG\cite{PDG2k}, namely $\eta = 0.358$. We then consider 
three cases:
\begin{itemize}
\item 
Fit1: All parameters $A, B$ and $\epsilon$ are left free 
\item 
Fit2: $\epsilon$ is fixed to 0.093, as measured in 
$pp$ and $\bar p p$ collisions, the other parameters are left free 
\item
Fit3: $\epsilon$ is fixed to 0.093, but a second pomeron term of the form 
$C s_{\gamgam}^{\epsilon_1}$ as proposed in~\cite{landshof}   
was added with $\epsilon_1= 0.418$ and the normalization $(C)$ fitted.
\end{itemize}
The fits are made for the L3 data alone -- which have the largest range--
and the L3+OPAL data. In the latter case the results remain dominated by
the L3 data, but the relatively low increase in $\chi^2$ value shows that 
these data sets are now well compatible with each other.
The results of the fit are shown in Fig.~\ref{figfit}.
Clearly the L3  data reject Fit2. 
A universal pomeron slope  of about 0.08-0.093
is not compatible with the L3 data. 
Fit1 shows that the pomeron term is of order 0.2-0.3.
This remains true if the fits are made
to the data corrected with either PHOJET or PYTHIA separately as the
L3 collaboration reports in their paper~\cite{L3}.
Using two pomerons terms with a fixed power following~\cite{landshof}, as 
in in Fit3, 
also accounts for the rise. Based on these data alone, and without taking
into account correlations between the data points, the second pomeron component
in the data is visible with about 5$\sigma$ significance.
These results do not change significantly if the data-point at the largest
$\sqrt{s_{\gamma\gamma}}$ value is excluded from the fit. The results do 
show a dependence on the model used for unfolding the cross-sections. We 
find for the L3 data, that correcting the data using Pythia or Phojet at a 
time, instead of taking an average of both these models, yields values of 
$\epsilon$ of $0.29 \pm 0.03$ and $ 0.20 \pm 0.03$ respectively for Fit1.
A similar model dependence in the fits was observed in ~\cite{blockkang}.
Note however, that both values of $\epsilon$ are still significantly 
larger than the soft pomeron value of $0.093$. Thus the total 
$\gamma\gamma$ cross section appears to rise faster than in hadron-hadron 
collisions. It is however imperative that other LEP experiments
make similar  analyses in the high $\sqrt{s_{\gamma\gamma}}$ region to
confirm this important result.

The results depend on the value of $\eta$ used in the fits.
E.g. changing $\eta$ in the range of $0.3-0.45$ yields a slope 
$\epsilon $  of $0.280-0.215$ in Fit1 and the cross section for the
second hard pomeron from $5.9\pm 1.2 - 4.6\pm1.0$ in Fit3.
While the main message of the fits to the data remains the same for these
values, the exact value of the pomeron slope or cross section depends on 
the Regge power. To pin down the Regge part, better measurements at low 
energy in dedicated experiments will be needed.

Thus a study of high energy behaviour of the two-photon total cross section
holds potential of yielding very interesting information. Improved or 
additional data in the LEP range from other experiments can help but 
there are a number of theoretical and experimental issues, which only 
an $e^+e^-$  Linear Collider (LC) can clarify, by  reaching  higher energies.

We now turn to discuss the models proposed so far for 
$\sigma_{\gamma \gamma}$. A
review of the theoretical ideas and issues  on the total cross section
behaviour in $\gamma\gamma$ collisions was presented in~\cite{pancheri}.
As mentioned already, all models predict a rise of the cross-section with the 
collision energy \syy ,  but with very different slopes.  
The dramatic differences in the predictions for high energies show 
our present lack of understanding.  In pure proton-like models
(for example the solid curve~\cite{aspen}), the rise 
follows  closely that of the proton-proton cross-section, while in  QCD based 
models (upper~\cite{BKS} and lower~\cite{pancheri} bands), the rise is 
obtained using the eikonalized pQCD jet cross-section.  The upper and lower 
here refer to the position of the bands at low energy. In order to give a
quantitative estimate of the  energy dependence in the different models, one
can fit the model predictions with  eq.\ref{eq1} and
calculate, for each model,  the effective values of  $\eta$ and $\epsilon$. The
results for three of the models   are given in  
\begin{table}[htb]
\begin{center}
\caption{Values of the parameters $A, B, \eta, \epsilon$ of Eq.~\ref{eq1}
obtained by numerical fits to the various model predictions. The upper and
lower refer to the position of the edge at high energies.}
\vspace{0.3cm}
\begin{tabular}{|c||c|c|c|c||}
\hline
Model & $A$ (nb) & $\epsilon$ & $B$ (nb) & $\eta$ \\ \hline
\hline
BKKS (upper edge)  & 166.5  & 0.13 & 538.2 &  0.38 \\ 
BKKS (lower edge)  & 180.6 & 0.11 & 356.5 &  0.18  \\ 
Aspen  & 145.7 & 0.094 & 517.5 &  0.39   \\  \hline
EMM (lower edge) &14.01  &0.34  &475.4 & 0.14    \\  
EMM (upper edge) & 19.9  &0.29  &475.3 & 0.084   \\  \hline
\end{tabular}
\end{center}
\label{table3}
\end{table}
Table~\ref{table3}.  As
expected, the slope for the Aspen model is close to that of the proton 
cross-sections, while the QCD based models, EMM or
BKKS, are described by $\epsilon$ values in the range $0.11$--$0.34$.
Note, however, that the earlier discussion of the fits  shows that the lower 
values of $\epsilon$ are in disagreement with the present data.

The discussion above just shows  that data seem to indicate larger values for 
$\epsilon$ in eq.  (1) for $\gamma \gamma$ processes, i.e. a faster rise with 
energy than for the $pp/p\bar p$ processes.  In addition, we also see that  
different models differ greatly in their predictions.  All this makes it clear 
that a good measurement of the total hadronic cross-section for photon 
induced processes is quite important.  In the next sections we will examine
the experimental  issues involved in measuring two-photon interactions at a
future high energy  \ep\ and \gamgam\  collider, extending previous
estimates up to energy of 1 TeV in the $\gamma \gamma$ centre of mass frame.

\section{Linear colliders and Two Photon  Physics.}
Two photon processes at future high energy  LC can be
measured either as in a storage ring, via photon emission from the 
lepton beams, according to a  Weizs\"acker Williams(WW){\cite{weisz}
 energy distribution, or using the 
LC in a photon collider mode~\cite{telnov}.
In the latter case the
high energy electron beam is converted into a high energy photon
beam, by backscattering of photons off  an intense laser beam, just before
the interaction point. The maximum energy of the generated photons is
given by $E_\gamma^{max} = xE_e/(1+x)$, with $E_e$ the electron beam energy
and $x = 4E_eE_L\cos^2(\theta/2)/m_e^2c^4$ with $E_L$ and $\theta$ the laser
photon energy and angle between the electron and laser beam.
The distance of the conversion to the interaction point is in the range
of several mm to a few cm. A typical value for $x$ is 4.8, which leads
to photon spectra which peak around $0.8E_e$.  Hence, a typical distribution 
of $\gamma\gamma$ luminosity as a function of the invariant mass peaks at 
the maximum reachable invariant mass of around $0.8\sqrt{s_{e^+e^-}}$
 with a width of $\approx 0.10-0.15$ for $\gamma\gamma$ collisions. The 
'luminosity' is usually defined to be the luminosity corresponding to the  
region $0.8 \sqrt{s_{\gamma\gamma, max}}  < \sqrt{s_{\gamma\gamma}}
< \sqrt{s_{\gamma\gamma, max}}$ and 
is typically 10\% of the geometrical $e^+e^-$ luminosity.
However, since the electron beams are converted
before collision, smaller $\beta^*$ functions at the interaction point
can be allowed for. This can make up for part of the luminosity lost 
compared to an  $e^+e^-$ collider, by using smaller beam spots. 
For  TESLA~\cite{tesla}, for an LC with initial CMS energy of 500 GeV, one 
finds~\cite{telnov2} ${\cal L}_{\gamma\gamma} \simeq 0.35 {\cal L}_{e^+e^-}$
This leads to event samples corresponding to a luminosity of the order 
100 fb$^{-1}$ per year, for the Photon Collider. Similar numbers are obtained 
for CLIC~\cite{clic}.

In this study the PHOJET~\cite{phojet} program was used which has an option 
for a simple backscattered laser (BL) photon spectrum. Both the \ep and BL
modes of the LC were investigated. Detector effects were 
studied with the program SIMDET~\cite{simdet}.

\section{Simulation study}

First we consider the case for the \ep collider mode.  Present data for the 
two photon cross-section at LEP are hampered by  experimental and detector 
limitations. For two photon events coming from interactions of quasi-real 
photons, the electrons disappear in the beam-pipe. Hence the only information 
available is the hadronic final state.  The variable \syy \ needs to be 
reconstructed from the visible hadronic final state in the detector.  At the 
highest energies for a 500 GeV $e^+e^-$ LC the hadronic final state extends in 
pseudorapidity $\eta = \ln\tan\theta/2$ in the region $-8 < \eta < 8$, as 
shown in Fig.~3a, while a typical LC detector covers roughly  the region 
$-3 < \eta <3$. Hence the correlation of the measured \syy \ compared to the 
true one will be even poorer at the LC than at LEP, as shown in Fig.~3b.
However, some information can be obtained by measuring the total integrated
cross-section above a value \syy (see Section 5).

For a $\gamma \gamma$ collider the photon 
beam energy can be tuned with a spread of less than 10\%, such that
measurements
of \sigg $\,$ can 
be made at a number of ``fixed'' energy values in e.g. the range
 $100 < $~\syy~$ < 400$ GeV by changing
the beam energy of the collider, as shown in Fig.~3c. For the
spectra shown
it was assumed in the simulation  that the same value of $x$ is kept at each
energy, which  means that the 
wavelength of the laser needs to be changed
at each beam energy. A more likely scenario is that the same laser is used 
at the lower energies, which will then have correspondingly different 
$x$ values, and thus somewhat different spectra.

In the simulation study events were selected with visible 
$\sqrt{s_{\gamma\gamma}} > 10 $ GeV, 
and it was asumed that a minimum of three charged tracks would be detected in 
the experiment. The acceptance for non-diffractive events was found to be  
about 95\%. The simulation shows that the diffractive non-elastic events 
are accepted with 35\% efficiency, while elastic events are essentially lost.
Hence to measure the total cross-section with sufficient precision 
specially designed measurements for the diffractive components will
be required. To that end also measurements at lower energy 
(e.g. LEP) will be useful, in order to extrapolate the elastic component.
A technique to measure diffractive contributions separately, mirrored to 
the rapidity gap methods used at HERA, has been proposed in \cite{engel}.
It is assumed here that such measurements will be made, and it will be 
possible to reduce the systematic error due to  the diffractive component.
\begin{figure}[htb]
\begin{center}
\epsfig{file=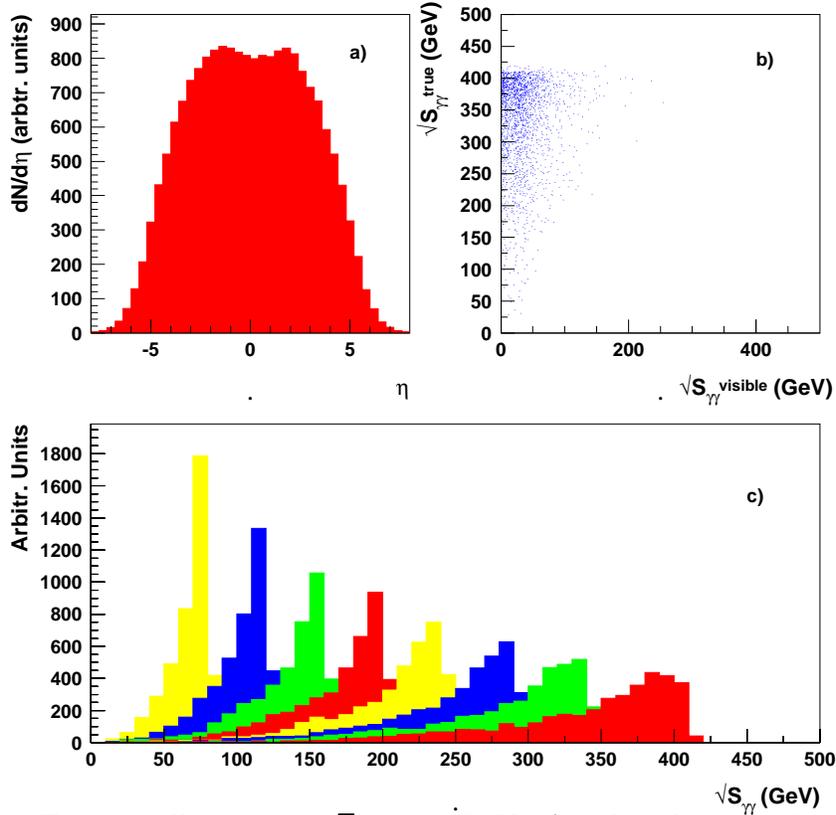,bbllx=80,bblly=180,bburx=470,bbury=620,width=8.5cm}
\caption{For $\gamma\gamma$ 
collisions at $\sqrt{s}$ = 400 GeV:
a) $\eta$ distribution of produced particles; b) correlation between the 
true and visible $\sqrt{s}$; c) luminosity spectra for eight
measurements of $\sigma_{tot}$ at different energies.}
\label{corrections}
\end{center}
\end{figure}
\begin{table}[htb]
\begin{center}
\caption{
List of systematic errors for the total cross-section measurement.
The third column comments on the source/reason }
\vspace{0.3cm}
\begin{tabular}{|c||c|c||}
\hline 
 Type & Value &  Comment \\ \hline
\hline
Selection cuts    & 3\%  &$\sim$HERA/frag. parameters  \\ 
Diffraction    & 3-8\% & $\rho\rho$ events  \\ 
Detector smearing  & 3\% & $\sim$HERA    \\  \hline
Lumi/$E_{\gamma}$ spectrum   & 3-4\%& short runs ($\sim$ one day)  \\ 
Bin correction   & 2\%& $\sqrt{s_{\gamma\gamma}}$  spread  \\ \hline
\end{tabular}
\label{table0}
\end{center}
\end{table}
The absolute precision with which
these cross-sections can be measured ranges from 5\% to 10\% for 
collider like TESLA, in the CMS energy range up to 700 GeV. 
The important  contributions to the errors come from 
the control of the diffractive component  of the cross-section,
Monte Carlo models used to correct for the event selection cuts,
the knowledge of absolute luminosity and  shape of the luminosity spectrum.  
For a photon collider based on CLIC~\cite{clic}, which 
could reach $\sqrt{s_{\gamma\gamma}}$
values  up to 2 TeV and beyond,  the corresponding error
range is from about 7 to 15\%.
The errors  given in Table.~\ref{table0} are  estimated
using SIMDET~\cite{simdet} and extrapolations from HERA and LEP
studies.

\begin{figure}[htb]
\begin{center}
\epsfig{file=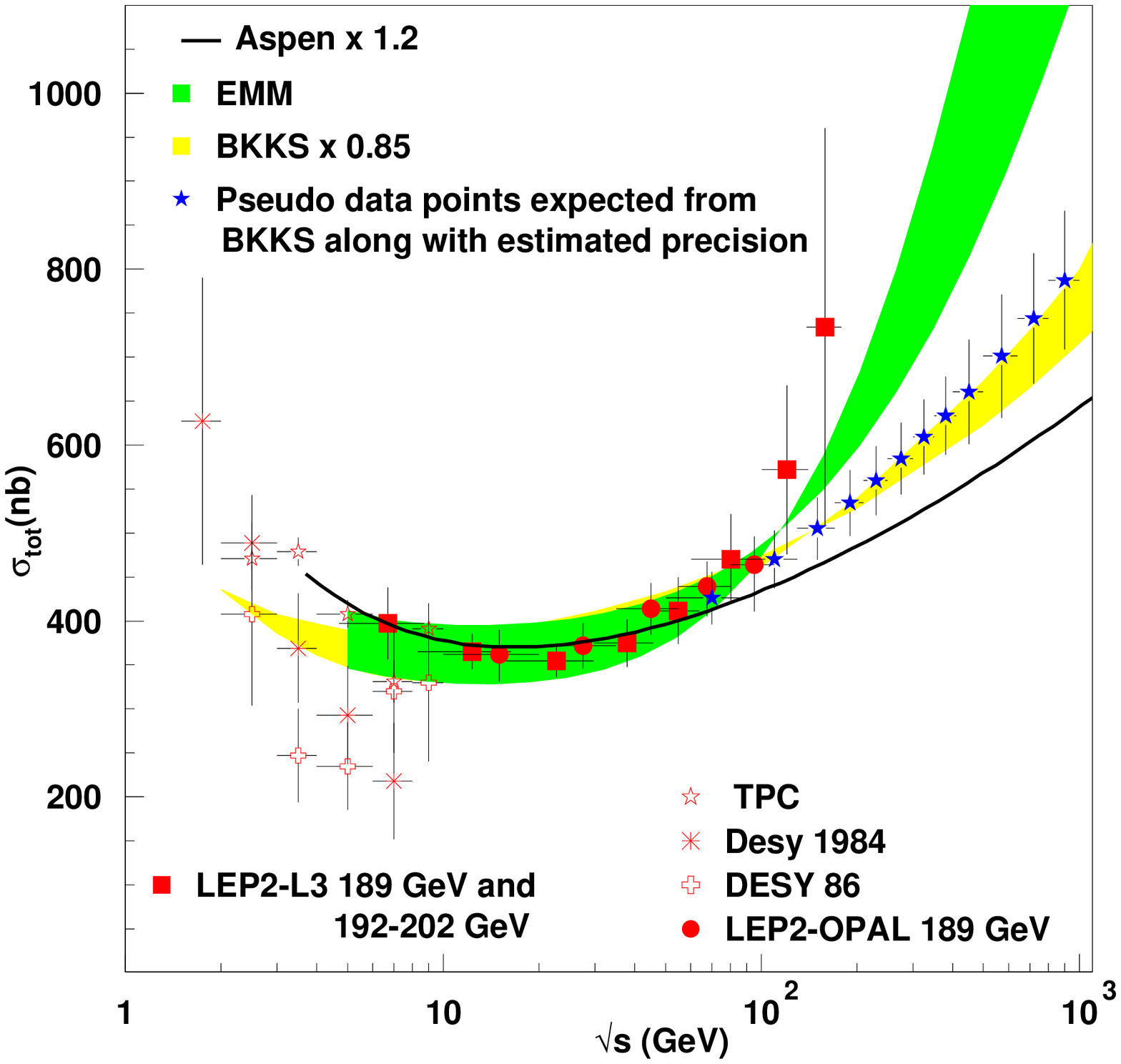,bbllx=50,bblly=160,bburx=500,bbury=580,width=8.cm}
\caption{
The total $\gamma\gamma$  cross-section as function of the collision energy, 
compared with model calculations:
BKKS band (upper and lower limit correspond to different   
photon densities \protect \cite{BKS});
a proton-like model (solid line \protect \cite{aspen});
EMM band (Eikonal Minijet Model for total and 
inelastic cross-section, with different photon densities and 
different minimum jet transverse momentum \protect
\cite{pancheri}). The proton-like and BKS
models have been normalized to the data, in order to show the 
energy dependence of the cross-section.}
\label{six2}
\end{center}
\end{figure}

Fig.~\ref{six2} shows present photon-photon cross-section data
in comparison with predictions at higher photon energies 
from some of the recent phenomenological 
models~\cite{pancheri} illustrated in Fig.~1. 
Because of uncertainties in the overall
normalization, some of the  curves reproduced in this figure
 have been  scaled (up or down) so as to  overlap with the low energy data
 point from LEP experiments. This emphasizes  solely the energy dependence 
and reduces the uncertainties due to the low energy normalization. 
Pseudo data points from one of (thus scaled) models \cite{BKS}
are added with a systematic error of $7\%$--$11\%$. The statistical 
errors are small since even only for a day of
running at a given energy, $O( 10^{7})$ events are produced.
Statistical samples of order $10^5$ events will be sufficient
for this measurement.

\section{Precision needed to distinguish between \mbox{models}}
In this section  we update previous estimates \cite{pancheri,ismd99}
for the precision needed to distinguish between different models for 
the total $\gamma \gamma$ cross-section. 
\begin{table}[bth]
\begin{center}
\caption{Expected \gamgam\ cross-sections and precision required for 
in their  measurement to distinguish between the different `proton' like 
models}
\vspace{0.3cm}
\begin{tabular}{|c||c|c|c|c||}
\hline 
$\sqrt{s_{\gamma \gamma}}$ (GeV) & Aspen &  BSW & DL & $1 \sigma$ \\ \hline
\hline
 20    & 309 nb & 330 nb & 354 nb &  7\%  \\ 
 50    & 330 nb & 368 nb & 402 nb &  10\%   \\ 
 100   & 362 nb & 401 nb & 450 nb &  10\%   \\  
 200   & 404 nb & 441 nb & 507 nb &  9\%   \\  
 500   & 474 nb & 515 nb & 598 nb &  8\%   \\  
 700   & 503 nb & 543 nb & 636 nb &  8\%   \\ 
1000   & 538 nb & 578 nb & 679 nb &  7\%   \\ \hline
\end{tabular}
\label{table1}
\end{center}
\end{table}
In Table ~\ref{table1} we show total $\gamma \gamma$ cross-sections for three
models of the `photon-is-like-the-proton' type. The last column shows 
the 1$\sigma$ level precision needed to discriminate between Aspen\cite{aspen} 
and BSW\cite{ttwu} models. The model labelled DL is obtained from
Regge/Pomeron exchange and factorization at the residues \cite{DL}, 
with parameters from ref.\cite{PDG2k}.  Table 5 gives the precision needed 
to distinguish   between the two minijet 
formulations of Fig.~\ref{six} and the BKKS model\cite{BKS}.
\begin{table}[bth]
\begin{center}
\caption{Expected \gamgam\ cross-sections and precision required in their 
measurement in order to distinguish between different formulations of 
the EMM and BKKS~\protect\cite{BKS}, whose common characteristic is that  
the high energy rise is computed with a QCD input.}
\vspace{0.3cm}
\begin{tabular}{|c||c|c|c|c||}
\hline 
$\sqrt{s_{\gamma \gamma}} (GeV)$ &EMM, Inel,GRS &EMM, Tot,GRV & BKKS& $1 \sigma$ \\ 
& ($p_{\rm tmin}$=1.5 GeV)& ($p_{\rm tmin}$=2 GeV)              & GRV & \\ \hline
\hline
 20    &399  nb & 331 nb      & 408 nb &   2 \%  \\ 
 50    &429  nb & 374 nb      & 471 nb &   9\%   \\ 
 100   &486  nb & 472 nb      & 543 nb &   3\%   \\  
 200   & 596 nb & 676 nb      & 635 nb&   6\%   \\  
 500   & 850 nb & 1165 nb     & 792 nb &  7  \%   \\  
 700   & 978 nb & 1407 nb     & 860 nb &   13 \%   \\ 
1000   & 1133 nb& 1694 nb     & 940 nb & 19\%   \\ \hline
\end{tabular}
\label{table2}
\end{center}
\end{table} 
The last column in Table~\ref{table2} gives the percentage 
difference between the two models which bear closest results among these
three.

A detailed comparison of the 
predictions reveals that in order to distinguish between 
all the models the cross-sections  need to be determined 
to a precision of better than 10\%~\cite{pancheri} 
at a future 0.5-1 TeV \ep collider.
This seems feasible from the simulation studies discussed above. It should
be pointed out that some of the differences  in the model predictions even 
at higher energies are  due to the low energy normalization. If all the models 
were to be scaled down to the same  value, for instance at 50 GeV c.m. energy, 
then differences among the different proton like models would very much 
decrease.  Not so for the QCD based models, which continue to  show 
very different high energy behaviour.

On the other hand, 
while the absolute cross-sections are measured with limited precision, the 
change of the cross-section with energy can be determined much more 
accurately. Fitting the  data of the collider to the Regge inspired form 
$s^{\epsilon}$ in the 
high energy region, one can determine $\epsilon $ with a precision 
of $\Delta \epsilon = 0.02$. The models show a variation between 
$\epsilon = 0.08$ and $\epsilon = 0.26$ and hence can be distinguished.

\section{$\sigma_{tot}$ in the $e^+e^-$ mode}

It will be difficult to measure $\sigma_{tot}$ with sufficient
precision in the normal $e^+e^-$ mode of the linear collider,
however some information can be gained from measuring the 
cross-section $\sigma(e^+e^- \rightarrow e^+e^- {\rm hadrons})$.
 This is calculated by convoluting 
the \gamgam\  total  cross-section with the spectrum  of these photons.
This spectrum is given by the  Weizs\"acker Williams(WW) or effective 
photon approximation\cite{weisz} which has been used quite successfully 
to translate photoproduction cross-sections into those for electron initiated
processes. There have been many discussions of the improvements on the 
original WW approximation~\cite{newww}. This has also been extended to 
include the effects of a reduction in the parton content of the photon 
due to virtuality of the photon~\cite{manmev}, while dealing with the
resolved photon processes. The cross-section, including the effects 
due to (anti)tagging of the electron is given by
\begin{equation}
\sigma^{\rm had}_{\eplem} = \int_{zmin}^1 dz_1 
\int_{zmin/z_1}^1 dz_2 \; f_{\gamma/e}(z_1) f_{\gamma/e}(z_2) 
\sigma(\gamgam \to {\rm hadrons}).
\label{ggtoee}
\end{equation}
Here $z_{min} = s_{min}/s$ where $\sqrt{s}$ is the c.m. energy of the \eplem\  
collider. The WW spectrum used is given by
\be
f_{\gamma/e} (z) = {{\alpha_{\rm em}}\over {2 \pi z}} 
\left[ (1 + (1-z)^2) \ln {{P^2_{max}}\over {P^2_{min}}} -2(1-z) \right],
\label{wwtag}
\ee
where $$
P^2_{max} = s/2*(1-\cos \theta_{tag}) (1-z), P^2_{min}= m_e^2 {z^2 \over (1-z)}.
$$
Here, using $\theta_{tag}$  the maximal scattering angle for the outgoing 
electron, we have taken anti-tagging into account and have included the  
suppression of the photonic parton densities due to its virtuality following
ref.~\cite{zpcus}. 

To select $e^+e^- \rightarrow e^+e^- {\rm hadrons}$ events, a minimum
value  of $s_{\gamma\gamma}$ is required, selecting a region such that the 
value of $s_{\gamma \gamma}$  can be corrected for smearing and losses with 
sufficient precision. Also a maximum value is imposed, because the events 
resemble annihilation events for too large a value of  $s_{\gamma\gamma}$
and cannot be easily separated. 
Additionally an anti-tagging condition for the scattered electrons is imposed. 
SIMDET simulation studies lead to choose the region
50 GeV$^2$ $ < s_{\gamma\gamma}< 0.64 s_{ee}$, and the anti-tagging cuts
are $\theta_{\rm tag} =  0.025 , E_{min}^e = 0.2E_{\rm beam}$.
With these cuts the total cross-section can be determined with a 
precision of 5-10\%.

In Fig.~\ref{fig4}, we  show the cross-section 
\begin{figure}
\begin{center}
\epsfig{file=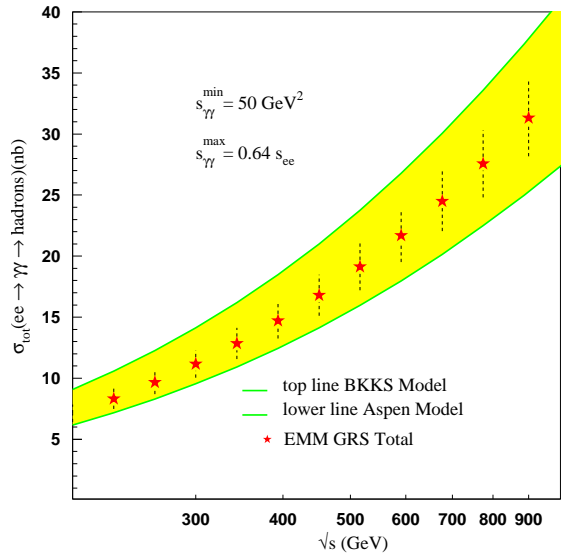,bbllx=40,bblly=160,bburx=520,bbury=620,width=7cm}
\caption{Cross-sections for hadron production  due to \gamgam\
interactions in  \eplem\  reactions.}
\label{fig4}
\end{center}
\end{figure}  
as a function of $\sqrt{s}$ of the  \eplem\ machine.  The top curve 
corresponds to the prediction  for $\sigma_{\gamgam}$ of the BKKS model
\cite{BKS}  
and the lower curve corresponds to the prediction of the Aspen
model~\cite{aspen}. Note that the difference 
of about factor 2 (say) at $\rts = 700$  GeV, is reduced to about 
$30-40\%$ after convolution with the bremsstrahlung spectrum. 
The `data points' are given with a pessimistic error of 10\%.
Hence even in the worst of the cases (no photon collider, large uncertainties 
in the $e^+e^-$ $\sigma^{tot}_{\gamma\gamma}$ cross-section measurement) a 
linear collider with $\sqrt{s_{e^+e^-}} = 500$ GeV, can definitively 
contribute to a further understanding of the energy dependence of the total 
$\gamma\gamma$ cross-section.

Eq.~\ref{ggtoee} can also be used to calculate the number of 
hadronic events per bunch crossing, which  is expected to be
significant at the high energy $e^+e^-$ collider like CLIC. However, in this
case it is necessary to add the effect of the beamstrahlung photons as well. 
We do this for CLIC, by taking the spectra of beamstrahulng photons
as provided in Ref. \cite{schulte}.  For the design parameters considered,
the two-photon luminosities per bunch crossing,  corresponding to both or one 
photon being a bremsstrahlung one are  
${\cal L}_{ee}^{\gamma \gamma}=6.4852 \times 10^{34} m^{-2}$ and 
${\cal L}_{eg}^{\gamma \gamma}= 5.3589 \times  10^{34}m^{-2}$ respectively, 
where as the one coming from beamstrahlung photons alone is  
${\cal L}_{gg}^{\gamma \gamma } = 4.9534  \times 10^{34}m^{-2}$.  The 
expected number of hadronic events expected per bunch crossing, with 
these effective two photon luminosities per bunch crossing, are shown 
in Table~\ref{tab:tablebs}. It shows the number of events expected for 
three different values of the lower limit on $s_{\gamma \gamma}$ instead 
of the fixed value of 50 GeV$^2$ considered above.

\begin{table}
\begin{center}
\caption{\label{tab:tablebs}Number of events per bunch crossing expected at CLIC.}
\vspace{0.3cm}
\begin{tabular}{|c||c|c|c|c||}
\hline
$s_{min}$~~ GeV$^2$&Aspen &EMM(BN)&BKKS&EMM \\ \hline
\hline
 5   &4.  &5.5  &5.7  &6.3    \\  
25  &3.4&4.7  &5.0  &5.5     \\
50  &3.2&4.5  &4.7  & 5.3    \\
\hline
\end{tabular}
\end{center}
\end{table}

The number obtained by us for the Aspen model is consistent with that in 
Ref.\cite{daniel} with SAS parametrisation of the 
$\sigma_{\rm tot}(\gamma \gamma \rightarrow {\rm hadrons})$.  Thus we see that 
depending on which theoretical model gives the right high energy description,
we expect  between 4-7 hadronic events per bunch crossing at CLIC.  
The beamstrahlung photons completely dominate the 
$\gamma \gamma $ luminosity. Inclusion of the beamstrahlung contribution 
increases the expected number of events by about a factor 10 than expected 
just for the bremsstrahlung photons.  However, about half of these events 
come from the contribution to the $\gamma \gamma$ luminosity coming from the 
cross-term between the bremsstrahlung and the beamstrahlung photons.

\section{Conclusions}

Future linear $e^+e^-$ colliders will be instrumental in the study of 
high energy photon collisions. We have shown that it will be possible at 
these colliders, to extract new important information on the energy 
dependence of the total hadronic cross-section in two photon collisions, 
$\sigma^{\rm had}_{\gamgam}$.  The best option for this purpose 
would be the photon collider mode, allowing for 
precise  measurements at several $\sqrt{s_{\gamma\gamma}}$ 
energies. Accuracies of the order of 5-10\% can be achieved, in a region
where model predictions vary by a factor 2 or more. Even in  the $e^+e^-$ 
mode, there is sufficient sensitivity left in the total inclusive reaction 
$e^+e^- \rightarrow e^+e^- {\rm hadrons}$ (which does not require 
reconstruction of \syy), to distinguish between model classes though  
perhaps not between models in a given class. Further, it should be added 
that, for centre of mass energies of the $e^+e^-$ collider up to a 
$800$ --$1000$ GeV, the size 
of this inclusive cross-section is determined more by the normalisation of 
$\sigma (\gamma \gamma \rightarrow {\rm hadrons})$ for \syy\ in $100$ -- $200$ 
GeV region than by the steepness of its energy dependence.  Thus clarification 
of the measurements by the LEP groups in this energy range can  play a crucial 
role in the determination of the hadronic backgrounds per bunch crossing 
expected in the $e^+e^-$ mode.

\section*{Acknowledgments}
This work was supported in part by the EU through HPRN-CT2002-00311.
RMG wishes to thank the CERN theory division for hospitality where part of 
this work was done and the Department of Science and Technology, India 
for partial support under project no. SP/S2/K-01/2000-II.

\end{document}